\documentclass[floatfix,nofootinbib,prd,amsfonts,twocolumn]{revtex4-1}

\usepackage{graphicx}
\usepackage{epic}
\usepackage{eepic}
\usepackage{latexsym}
\usepackage{amssymb,amsmath}

\newcommand{\eq}[1]{(\ref{#1})}
\newcommand{\be}{\begin{equation}}
\newcommand{\ee}{\end{equation}}
\newcommand{\bea}{\begin{eqnarray}}
\newcommand{\eea}{\end{eqnarray}}

\newcommand{\hs}[1]{\hspace{#1 mm}}

\newcommand{\tf}{\tilde{\phi}}

\def\a{\alpha}

\def\C{\Gamma}
\def\d{\delta}
\def\D{\Delta}
\def\e{\epsilon}

\def\f{\phi}
\def\fr{\frac}

\def\k{\kappa}
\def\l{\lambda}
\def\L{\Lambda}
\def\m{\mu}

\def\th{\theta}

\def\del{\partial}

\let\bm=\bibitem
\def\nn{\nonumber}

\begin{document}

\title{Exact RG Flow  in an Expanding Universe and Screening of  the Cosmological Constant} 

\author{Ali Kaya}
\email[On sabbatical leave of absence from Bo\~{g}azi\c{c}i University. \\]{ali.kaya@boun.edu.tr}
\affiliation{Department of Physics, McGill University, Montreal, QC, H3A 2T8, Canada\\
and\\
Bo\~{g}azi\c{c}i University, Department of Physics, 34342, Bebek, \.{I}stanbul, Turkey}

\begin{abstract}
We obtain the exact renormalization group (RG) flow equation for a self-interacting real scalar field in an expanding cosmological background. The beta functional for the potential in the local potential approximation is determined in terms of the mode functions defining the vacuum of the theory. For a scalar in an inflating universe, which is released in the Bunch-Davies vacuum, the flow equation is analyzed in detail using the optimized cutoff function for the quartically truncated potential. In that case, the constant term in the potential, which can in fact be viewed as the cosmological constant, decreases as the IR cutoff is removed, supporting the existence of a quantum mechanical screening mechanism. On the other hand, the running of the quartic coupling constant at IR is altered by the expansion, which seems to affect the quantum triviality of the theory.  
\end{abstract}

\maketitle

\section{Introduction}

The exact renormalization group (ERG) method is an important tool that can be used to probe nonperturbative properties of quantum field theories (for review see e.g. \cite{rev1,rev2,rev3,rev4,rev5,rev6}). In the form derived in \cite{wett},  the ERG flow gives an interpolation between the UV (or fundamental) action describing the microphysics and the full (IR) quantum effective action, which encodes the macroscopic phenomena. The method offers an effective way of calculating the beta functions and the nontrivial fixed points of RG flows. The ERG is obtained by introducing an IR cutoff function in momentum space, which is arbitrary as long as it satisfies certain conditions. Although the physical results become independent of the cutoff function after the regulator is removed, some specific choices are known to optimize the convergence and the stability of the flow \cite{lit1,lit2}. In general, the flow equation is an infinite-dimensional functional differential equation, which can be simplified in various ways and/or truncations. For instance, in the local potential approximation (LPA), which is first introduced in \cite{loc1,loc2}, one may take a momentum-independent limit of the ERG with constant field values. It is also possible to employ a systematic derivative expansion \cite{mor} (or expansions in powers of fields or canonical dimensions, see e.g. \cite{rev1}). The standard perturbation theory can be recovered from the exact flow by an iteration in the coupling constant \cite{lit4,lit5}. 

The ERG techniques are applied to pure gravity and gravity with matter systems in searching for nontrivial UV fixed points implying asymptotic safety \cite{w1}. There are also suggestions that the long-distance gravity is described by an IR fixed point \cite{ir1,ir2,ir3,ir4} (see also \cite{o1,o2,o3,o4,o5,o6,o7} which discuss possible implications of the running cosmological and Newton's constants). Recently, a concrete cosmological model is proposed, which takes into account the RG running of gravitational and matter coupling constants \cite{ergc1} (see also \cite{ergc2}, which takes the cutoff to be proportional to the Ricci scalar giving an effective RG improved $f(R)$ model ).  

In this paper, we consider a self-interacting real scalar field propagating in a fixed cosmological background and obtain the ERG flow equations for the in-out and in-in quantum effective actions. Our method differs from the above-mentioned papers in that we try to determine the impact  of the cosmic evolution on the quantum dynamics of the scalar field and ignore any possible backreaction effects. In other words, our approach fills the gap between the flat space studies and the works that include the gravitational dynamics. After introducing a cutoff function in the {\it comoving} momentum space that is suitable for a scalar in a cosmological background, the Lorentz signature ERG equations can be obtained both for the in-out and in-in formalisms. Unlike the flat space case, the ERG equations must be interpreted with care due to the explicit time dependence of the background. Following \cite{pol1}, we assume that self-interactions are turned on for a finite time affecting the running of the coupling constants. This interpretation is viable for inflation where one is interested in the form of the effective action at the end of the exponential expansion. 

Although we first derive exact flow equations using a comoving cutoff, it is also possible to obtain the ERG equations with a {\it physical} cutoff by redefining the flow parameter. Since the physical scale deviates from the comoving scale in time, the coarse graining in the effective action is different for both cases. While intuitively cutting off the physical scale seems the correct procedure, in the loop calculations the comoving IR cutoff can be argued to be  more natural \cite{ir-br}.  We will see that the two flows give qualitatively similar results in de Sitter space. 

As usual in the ERG analysis, one needs to employ some form of approximation to get information from the functional differential equation. The simplest route is to apply the LPA and analyze the flow for the effective quantum potential. In this approximation, we find that the beta functional for the potential is given in terms of the field mode functions defining the vacuum of the theory. We then consider an inflating universe and truncate the potential with a quartic polynomial after which the flow equations become a  system of  ordinary  nonlinear differential equations for three parameters; the cosmological constant, the scalar mass and the quartic coupling constant. The cosmological constant term in the potential is usually discarded in the flat space analyses but we keep it to understand its running with the renormalization scale. The flow equations become tractable in certain limits, which provides a qualitative understanding of the scale dependence of the physical parameters. 

The plan of the paper is as follows: In the next section we obtain the ERG flow equations both for the in-out and the in-in quantum effective actions. We then apply the LPA and determine the beta functional of the potential in terms of the field mode functions. In our approximation, the in-out and in-in cases become identical, which is similar to the one loop calculations of the quantum effective potential. In section \ref{3}, we focus on inflation and truncate the potential up to quartic order to study the runnings of the relevant parameters in detail. In the conclusions we summarize our findings and comment on some open problems for future work. 

\section{ERG flow in a cosmological spacetime}

In this paper we consider a self-interacting real scalar field $\phi$, propagating in a cosmological background that has the metric 
\be
ds^2=a(\eta)^2(-d\eta^2+d\vec{x}^2),
\ee
where $\eta$ is the conformal time. The scalar has the standard action:
\be\label{a}
S[\phi]=-\fr12\int d^4x \sqrt{-g} \left[\nabla_\mu\phi\nabla^\mu\phi+V(\phi)\right].
\ee
In flat space one usually writes the ERG equations in the Euclidean signature. Since it is not possible to Wick rotate a general cosmological spacetime, here one needs to work with Lorentz signature.

In obtaining the ERG flow, the action \eq{a} is modified by employing an IR cutoff function $R_\mu(k^2)$, which we take it to  depend on the {\it comoving} momentum variable $k$ and a {\it comoving} renormalization scale $\mu$. As in flat space, we require $R_\mu(k^2)$ to obey\footnote{As pointed out in \cite{rev1}, the third condition must ensure the divergence of the cutoff function $R_\m(k^2)$ in such a way that a delta functional arises in the path integral to kill all quantum fluctuations, see the discussion below.} 
\bea
&&R_\mu(k^2)\to 0\hs{2} \textrm{as}\hs{2}k\to\infty,\nn\\
&&R_\mu(k^2)\to \mu^2\hs{2} \textrm{as}\hs{2}k\to0, \label{con}\\
&&R_\mu(k^2)\to \infty\hs{2} \textrm{as}\hs{2}\mu\to\infty,\nn\\
&&R_\mu(k^2)\to 0\hs{2} \textrm{as}\hs{2}\mu\to0,\nn
\eea
and introduce the scale-dependent action as 
\be\label{ma}
S_\mu[\phi]=\fr12\int d\eta d^3k\, a^2\left(\tf'^2-[k^2+R_\mu(k^2)]\tf^2-a^2V\right),
\ee
where the prime denotes an $\eta$-derivative and $\tf$ is the Fourier transformed field in the momentum space
\be
\phi(\eta,\vec{x})=\fr{1}{(2\pi)^{3/2}}\int d^3 k\, e^{i\vec{k}.\vec{x}}\,\tf(\eta,\vec{k}).
\ee
We will use this tilde convention for all Fourier-transformed variables below. Since $\phi$ is real $\tf(\vec{k})^*=\tf(-\vec{k})$ and $k$ dependence of $\tf$ in \eq{ma} is suppressed. 

Let us introduce the {\it in-out} generating functional 
\be\label{iopi}
e^{iW_\mu[J]}=\int D\phi \exp\left(iS_\mu[\phi]+i\int d\eta d^3 x\, J\phi \right)
\ee
and the quantum effective action 
\be\label{lt}
\Gamma_\mu[\phi]=W_\mu[J]-\int d\eta d^3 x J\phi-\Delta S_\mu[\phi],
\ee
where $\Delta S_\mu$ is the extra cutoff-function-dependent piece added to the action \eq{ma}, i.e.
\be
\Delta S_\mu=-\fr12\int d\eta d^3k\, a^2\,R_\mu(k^2)\tf^2.
\ee
In general, to make the in-out path integral \eq{iopi} well defined, one must specify in and out asymptotic regions that would determine the necessary boundary conditions.  As an apparent manifestation of this problem,  one must actually fix the limits of $\eta$-integrals in \eq{iopi}. For the moment we ignore this important technical point by assuming that suitable conditions are imposed to make \eq{iopi} well defined and discuss the issue in more detail below.

Before writing down the ERG flow equation for the quantum effective action $\Gamma_\mu[\phi]$, let us briefly elaborate on the conditions \eq{con}. In flat space, each of these conditions is imposed  for a reason and one should make sure that they have the same consequences in this cosmological setup. While the first condition in \eq{con} ensures that the cutoff function does not change the UV behavior of the theory the last condition is needed to remove the cutoff completely, i.e.  $\Gamma_{\mu=0}[\phi]$ gives the full quantum effective action. 

The second condition in \eq{con} must provide that $R_\mu(k^2)$ works as an IR regulator. In flat space this is guaranteed by the fact that the corresponding term in the action actually becomes a mass term for IR modes. Although this is no longer true for the action \eq{ma}, since the mass term has a different geometric structure (note the scale factor dependencies), it is not difficult see that $R_\mu(k^2)$ functions as an IR regulator. Consider, for example, a massive scalar in de Sitter space. In the free theory, the mode function of the field is given by the Hankel function $\tf(k)=H_n(-k\eta)/\sqrt{-\pi\eta/4}$ where $n=\sqrt{9/4-m^2/H^2}$. Therefore, as $k\eta\to0$ one has  $\tf(k)\simeq 1/k^n$, and for a light field with real $n$, this singular behavior may give strong IR effects (or divergencies for the massless field). When the free action is modified as in \eq{ma}, the new mode functions can be obtained by $k^2\to k^2+R_\mu(k^2)$ and as $k\eta\to0$ one has $\tf(k)\simeq 1/\mu^n$ due to the second condition in \eq{con}. This\footnote{Note that the argument works for a comoving cutoff. If the cutoff is chosen to be physical then the mode equation is modified in a nontrivial way.} makes the theory IR safe. 

Finally, the third condition is imposed to satisfy $\Gamma_{\mu=\infty}[\phi]= S[\phi]$ so that $\Gamma_{\mu}[\phi]$ interpolates between the classical action and the full quantum action as $\mu$ runs from $\infty$ to 0. To verify this, one may use the functional integral representation of $\Gamma_{\mu}[\phi]$, which can easily be derived from \eq{iopi} and \eq{lt} to read
\bea
e^{i\Gamma_{\mu}[\phi]}=\int D\chi \exp\left(iS[\phi+\chi]-i\int \chi\fr{\delta\Gamma_\mu}{\delta\phi}\right.\nn\\\left.-\fr{i}{2}\int d\eta\, d^3 k\, a^2\,R_\mu(k^2)\,\tilde{\chi}(k)^2\right).
\eea
In the limit $\mu\to\infty$, the divergence of $R_\m(k^2)$ must ensure that the second line produces a delta functional
\footnote{To be more precise, the delta functional arises if $R_\mu(k^2)$ has a small imaginary piece. Otherwise one may argue that the integral becomes infinitely oscillatory which gives a nonzero contribution only when $\chi=0$.} 
\be\label{10}
\prod_k\d\left[\tilde{\chi}(k)\right],
\ee
which then implies $\Gamma_{\mu=\infty}[\phi]= S[\phi]$ by forcing $\chi=0$. For instance, similar to the $n$-dimensional identity $\lim_{\m\to\infty}\exp{(-\m^2 \sum_ix_i^2/2)}(\m/\sqrt{2\pi})^n=\prod_i\d(x_i)$, it is enough to have $R_\m(k^2)\to\m^2$ as $\m\to\infty$. Note that the delta functional is defined so that in a path integral $\int D\chi\d[\chi]F[\chi]=F[0]$ for any well-defined functional $F[\chi]$ and the extra constant terms arising from the delta functional limit can be absorbed in the path integral measure. We therefore see that the IR cutoff function introduced in the action as in \eq{ma} has all the desired properties. 

To derive the ERG flow equation, one may differentiate \eq{lt} with respect to $\mu$. The derivative of $W_\mu[J]$ can be found from \eq{iopi} as
\be
\fr{dW_\mu[\phi]}{d\mu}=-\fr12\int d\eta\, d^3k\,a^2\,\fr{dR_{\mu}}{d\mu}<\hs{-1}\tf^2\hs{-1}>.
\ee
On the other hand, the derivative of $\Delta S_\mu$ is 
\be
\fr{dS_\mu[\phi]}{d\mu}=-\fr12\int d\eta\, d^3k\,a^2\,\fr{dR_{\mu}}{d\mu}<\hs{-1}\tf\hs{-1}>^2,
\ee
where we used the fact that in the Legendre transformation \eq{lt}, the classical field $\phi$ equals the vacuum expectation value of the quantum field operator.  One thus obtains 
\be\label{e1}
\fr{d\C_\mu[\phi]}{d\mu}=\fr{i}{2}\int d\eta\, d^3k\,a^2\,\fr{dR_{\mu}}{d\mu}\fr{\d^2W_\mu[J]}{\d \tilde{J}(\eta,\vec{k})\d \tilde{J}(\eta,-\vec{k})},
\ee
where the source $J$ must be viewed as a functional of $\phi$. Note that the factor of $i$ in \eq{e1} arises since $<\hs{-1}\f^2\hs{-1}>-<\hs{-1}\f\hs{-1}>^2=-i\d^2W/\d J^2$. Since $\phi$, and thus $J$, are real, the Fourier-transformed variables are not independent. To avoid any complication, the functional derivatives in \eq{e1} can be defined as 
\be\label{un}
\fr{\d}{\d\tilde{J}(\eta,\vec{k})}=\fr{1}{(2\pi)^{3/2}}\int d^3 x\,e^{i\vec{k}.\vec{x}}\,\fr{\d}{\d J(\eta,\vec{x})},
\ee
which can be verified from the Fourier transformation and the chain rule. Therefore,  \eq{e1} can be rewritten as 
\bea
&&\fr{d\C_\mu}{d\mu}=\fr{-i}{2(2\pi)^3}\times\nn\\
&&\int d\eta\, d^3x d^3 y d^3k\,a^2\,e^{i\vec{k}.(\vec{x}-\vec{y})}
\fr{dR_{\mu}}{d\mu}G(\eta,\vec{x};\eta,\vec{y}), \label{erg1}
\eea
where the Green function is defined by 
\be\label{g1}
\int d\eta d^3 z \fr{\d^2(\C_\mu+\Delta S_\mu)}{\d\phi(\eta',\vec{x})\d\phi(\eta,\vec{z})}G(\eta,\vec{z};\eta'',\vec{y})=\d(\eta'-\eta'')\d(\vec{x}-\vec{y}).
\ee
Eq. \eq{erg1} is an exact functional differential equation for the quantum effective action $\C_\mu$, which gives ERG flow in a cosmological space. Note that in principle the Green function is determined by  $\C_\mu[\phi]$ from \eq{g1}. 

It is possible to repeat the above calculation for the in-in formalism. In that case the the generating functional is given by 
\be\label{iipi}
e^{iW_\mu[J^+,J^-]}=\int D\phi^\pm e^{i\left[S_\mu[\phi^+]-S_\mu[\phi^-]+\int \, (J^+\phi^+ - J^-\phi^-)\right]},
\ee
where $D\phi^\pm$ denotes the in-in path integral measure \cite{ak} and the modified actions are given by \eq{ma} both for $\phi^+$ and $\phi^-$. The quantum effective action is defined by the Legendre transformation
\bea
\C_\mu[\phi^+,\phi^-]&=&W_\mu[J^+,J^-]-\int \left(J^+\phi^+ - J^-\phi^-\right)\nn\\
&&-\left(\Delta S_\mu[\phi^+]-\Delta S_\mu[\phi^-]\right). \label{iil}
\eea
There is a finite upper time limit $\eta_f$ for the in-in path integral and the sum in \eq{iipi} is over all fields with $\phi^+(\eta_f)=\phi^-(\eta_f)$ (and $\phi'^+(\eta_f)=\phi'^-(\eta_f)$, see \cite{ak}). Therefore $\delta W_\mu/\d J^+(\eta_f)= -\delta W_\mu/\d J^-(\eta_f)$, which implies that the effective action $\C_\mu[\phi^+,\phi^-]$  is only defined for $\phi^+(\eta_f)=\phi^-(\eta_f)$. Differentiating \eq{iipi} with respect to $\mu$  and using \eq{iil} one may obtain
\bea
&&\fr{\C_\mu[\phi^+,\phi^-]}{d\mu}=\fr{i}{2}\int d\eta d^3 k\, a^2\,\fr{dR_{\mu}}{d\mu}\label{iif}\\
&&\times\left[\fr{\d^2W_\mu}{\d \tilde{J^+}(\eta,\vec{k})\d \tilde{J^+}(\eta,-\vec{k})}-\fr{\d^2W_\mu}{\d \tilde{J^-}(\eta,\vec{k})\d \tilde{J^-}(\eta,-\vec{k})}\right],\nn
\eea
where $J^\pm$ must be viewed as functionals of $\phi^\pm$ and the functional derivatives in the momentum space should be understood as in \eq{un}. Defining 
\bea
{\bf \tilde{\C}}(\eta,\vec{x};\eta',\vec{y})=\left[\begin{array}{cc} \fr{\d^2 \tilde{\C}_\mu}{\d J^+(\eta,\vec{x})\d J^+(\eta',\vec{y})} &  \fr{\d^2 \tilde{\C}_\mu}{\d J^+(\eta,\vec{x})\d J^-(\eta',\vec{y})} \\ \fr{\d^2 \tilde{\C}_\mu}{\d J^-(\eta,\vec{x})\d J^+(\eta',\vec{y})}&  \fr{\d^2 \tilde{\C}_\mu}{\d J^-(\eta,\vec{x})\d J^-(\eta',\vec{y})}
\end{array}\right],\nn
\eea
where 
\be\label{18}
\tilde{\C}_\mu=\C_\mu+\Delta S_\mu^+-\Delta S_\mu^-,
\ee
and the matrix Green function 
\be
\int d\eta d^3z {\bf \tilde{\C}}(\eta',\vec{x};\eta,\vec{z}){\cal I}{\bf G}(\eta,\vec{z};\eta'',\vec{y})={\bf I}\d(\eta'-\eta'')\d(\vec{x}-\vec{y}), \label{g2}
\ee
where
\be
{\cal I}=\left[\begin{array}{cc}1&0\\0&-1\end{array}\right],\hs{3} {\bf I}=\left[\begin{array}{cc}1&0\\0&1\end{array}\right],
\ee
the flow equation \eq{iif} can be written as
\bea
&&\fr{\C_\mu[\phi^+,\phi^-]}{d\mu}=\fr{-i}{2(2\pi)^3}\times\label{erg2}\\
&&\int d\eta\, d^3x d^3 yd^3k \,a^2\,e^{i\vec{k}.(\vec{x}-\vec{y})}\fr{dR_{\mu}}{d\mu} \textrm{Tr} \left[{\cal I}  {\bf G} (\eta,\vec{x};\eta,\vec{y})\right],\nn
\eea
where $\textrm{Tr}$ denotes the two-dimensional matrix trace. Eq. \eq{erg2} is the ERG flow equation for the in-in quantum effective action since the Green function ${\bf G}$ is in principle fixed by $\C_\mu[\phi^+,\phi^-]$ from  \eq{g2}. 

As mentioned above, the asymptotic boundary conditions must be specified to define in-out and in-in path integrals properly. However, once the ERG equation is obtained, one may forget about its origin and view it as  a property of the renormalized and finite quantum effective action. On the other hand, one still needs to impose certain boundary conditions to solve in-out and in-in Green functions, \eq{g1} and \eq{g2}, uniquely. As we will see, the boundary conditions are directly related to the vacuum chosen for the theory.  

From the first property in \eq{con}, the momentum integrals in \eq{erg1} and \eq{erg2} are guaranteed to converge at infinity (the three-dimensional spatial integrals correspond to Fourier transformations and thus they are not problematic). But $\eta$ integrals may diverge if the asymptotic regions and the  fall off conditions for the fields are not chosen carefully. To avoid any further complications one may follow \cite{pol1} and assume that the system is released at time $\eta_i$ in a specific vacuum state, which can be associated with a past infinity that is imagined to exist before $\eta_i$. For the in-in case this would fix a finite interval from this initial time $\eta_i$ to the final time $\eta_f$ at which the in-in correlation functions are calculated. Similarly, for the in-out case one may assume that the interactions (and the expansion of the space) are turned off at some time $\eta_f$, which can then be joined to a future asymptotic region. As a result we simply suppose a finite interval for the time integrals, which can be determined according to the physical problem at hand. For example, in applying the ERG for a scalar in an inflationary background, $\eta_i$ and $\eta_f$ can be taken to be the beginning and the ending of inflation. In that case the ERG gives how the exponential expansion affects quantum dynamics at a given renormalization scale. 

The quantum effective action can be systematically expanded in the number of field derivatives. For the in-out case, one may write 
\bea
\Gamma_\mu[\phi]&=&\fr12\int d\eta d^3x\,a^2\left[Z(\phi,\mu)(\phi'^2-\del\phi^2)-a^2V(\phi,\mu)\right]\nn\\
&&+...\label{qa1}
\eea
where the dotted terms represent fields with at least four derivatives. Note that the general covariance fixes the form of the two derivative terms, i.e. the wave-function renormalization $Z(\phi,\mu)$ is single. On the other hand, $Z(\phi,\mu)$ and the effective potential $V(\phi,\mu)$ are expected to depend on the conformal time explicitly since the background is time dependent. The kinetic operator can be calculated from \eq{qa1} as 
\bea
&&\fr{\d^2\C_\mu}{\d \phi(x)\d \phi(y)}=...+\label{k1}\\
&&\left[-a^2Z(\phi,\mu)\left(\fr{d^2}{d\eta^2}+\fr{2a'}{a}\fr{d}{d\eta}-\del^2\right)-\fr{a^4}{2}\fr{d^2V}{d\phi^2}\right]\d(x-y),\nn
\eea
where 
\be
x=(\eta,\vec{x}),\hs{5}y=(\eta',\vec{y}),
\ee
and the dots in \eq{k1} denote the terms containing at least two derivatives of the field $\phi$. Till now no approximation is used and to proceed one must calculate the Green function satisfying \eq{g1}. Obviously, the inverse of \eq{k1} is hard, if not impossible,  to calculate for arbitrary $\phi$. Indeed, even the operator that is first order in the derivative expansion, i.e. the second line of \eq{k1}, is difficult to invert\footnote{To invert \eq{k1} in the derivative expansion one may use the formula $(A+B)^{-1}=A^{-1}-A^{-1}BA^{-1}+...$ where $A$ represents the second line of \eq{k1} and $B$ corresponds to the higher-derivative terms indicated by the three dots.} for a generic $\phi$. However,  it is possible to evaluate \eq{erg1} at a constant field.  In that case, to  calculate \eq{g1} it is enough to consider the second line of \eq{k1}, which is very similar to the free kinetic operator other than the possible explicit conformal time dependencies of $Z(\phi,\mu)$ and $V(\phi,\mu)$. 

To find the Green function \eq{g1}, which is evaluated at a constant field $\phi$, one may  construct a free quantum operator as
\be
\hat{\phi}(x)=\fr{1}{a\sqrt{Z(\phi,\mu)}}\int\fr{d^3k}{(2\pi)^{3/2}}e^{i\vec{k}.\vec{x}}\chi_k(\eta)a_{\vec{k}}+H.c.
\ee
where $H.c.$ means Hermitian conjugate and the mode functions obey
\be\label{me}
\chi''_k+\left[k^2+\fr{R_\mu(k^2)}{Z(\phi,\mu)}-\fr{a''}{a}+\fr{a^2}{2Z(\phi,\mu)}\fr{d^2V}{d\phi^2}\right]\chi_k=0.
\ee
The free quantum operator $\hat{\phi}(x)$ should not be confused with the constant field $\phi$, which is used in the evaluation of  \eq{erg1}. With the help of $\hat{\phi}$, the Green function can be constructed as 
\be\label{ip}
G(x,y)=-i<T(\hat{\phi}(x)\hat{\phi}(y))>,
\ee
where $T$ denotes time ordering. Although the field operator $\hat{\phi}$ is introduced to calculate the Green function, there is a close connection with $\hat{\phi}$ and our original quantum scalar field $\phi$: In perturbation theory, the two fields actually coincide with each other at the lowest order. This allows us to fix the arbitrariness in the Green function, which shows up as the arbitrariness of the solutions of \eq{me}. Namely, in calculating \eq{ip}, one should use the solution of \eq{me} consistent with the vacuum chosen for the theory. Keeping this in mind, \eq{erg1} reduces to 
\be\label{pf}
\int d\eta\,a^4 \fr{dV(\phi,\mu)}{d\mu}=\fr{1}{8\pi^3}\int d\eta d^3k \fr{d R_{\mu}}{d\mu}\fr{|\chi_k(\eta)|^2}{Z(\phi,\mu)}, 
\ee
which can be viewed as the beta functional of the quantum effective potential. Let us remind the reader that in the flat space ERG calculations, a similar reasoning is used to find the exact propagator in momentum space where one simply takes the inverse of a polynomial obtained from the kinetic term of the effective action, which indeed refers to the vacuum. Since \eq{pf} is precisely  \eq{erg1} evaluated for a constant field, it is exact and free of any approximations. 

For the in-in case, the zeroth-order classical UV action is given by $\C[\phi^+,\phi^-]=S[\phi^+]-S[\phi^-]$. Although quantum corrections are expected to generate $\pm$ mixings in the effective action, we assume that 
\be\label{form}
\C_\mu[\phi^+,\phi^-]=\C_\mu[\phi^+]-\C_\mu[\phi^-],
\ee
where the functional $\C_\mu$  has the form \eq{qa1}. The main reason for this approximation is that the exact in-in effective action vanishes for $\phi^+=\phi^-$, i.e. $\C_\mu[\phi,\phi]=0$. Thus, the mixing terms can be expected to be suppressed (indeed in perturbation theory the mixing does not arise at one loop). As a crucial technical point, if one considers a general form for  $\C_\mu[\phi^+,\phi^-]$, then uniquely defining the matrix Green function \eq{g2} becomes a problem since one may no longer refer to the free theory. On the contrary, with the assumed form \eq{form} the operator ${\bf \C}$ defined above \eq{18} becomes similar to the free in-in kinetic operator and one may calculate the matrix Green function ${\bf G}$ by using the standard in-in formalism. In that case it is easy to see that the beta functional for the in-in effective potential, which is equal for $+$ and $-$ branches, becomes identical to \eq{pf}. Note that although one uses  the time-ordered and the anti time-ordered products in defining the diagonal entries of the in-in matrix Green function, this does not make any difference in the flow equation \eq{erg2} since the Green functions are evaluated at equal times. 

To proceed, we conveniently choose the optimized cutoff function \cite{lit1,lit2} 
\be\label{r}
R_\mu(k^2)=Z(\phi,\mu)(\m^2-k^2)\th(\mu^2-k^2).
\ee
It is easy to see that \eq{r} satisfies all conditions in \eq{con}. The presence of the wave function renormalization $Z(\phi,\mu)$ in \eq{r} is required to preserve the scaling behavior of the scalar field (see e.g. \cite{rev1}). Using \eq{r} in \eq{me} one sees that the mode equation becomes independent of $k$ for the regularized IR modes  
\be\label{me2}
k\leq \m:\hs{3}\chi''_\mu+\left[\mu^2-\fr{a''}{a}+\fr{a^2}{2Z(\phi,\mu)}\fr{d^2V}{d\phi^2}\right]\chi_\mu=0.
\ee
Thus  the momentum integral in \eq{pf} can be carried out to yield 
\be\label{pf2}
\int d\eta\,a^4 \fr{dV(\phi,\mu)}{d\mu}=\fr{\mu^4}{3\pi^2}\int d\eta  \left(1-\fr{\eta_\phi}{5}\right)|\chi_\mu(\eta)|^2,
\ee
where $\eta_\phi$ is the anomalous dimension
\be
\eta_\phi=-\fr{\del \ln Z(\phi,\mu)}{\del\ln \mu}.
\ee
Recall that in solving \eq{me2} one should refer to the vacuum chosen for the theory.

\subsection*{The Flow with a Physical Cutoff}

The cutoff parameter $\m$ refers to a comoving energy scale. If one would like to regulate {\it physical} IR modes then one should change $\m$ with time. In that case a new physical energy scale $\a$ can be introduced so that  
\be\label{mu}
\m=a(\eta)\,\a.
\ee
The cutoff function $R_\m(k^2)$ depends on $\a$ through \eq{mu} and it also changes with time. In this new setup, $\a$ can be used as the new flow parameter. 

It is possible to repeat the derivation of the beta functional, this time taking the $\a$ derivative of the quantum effective action in \eq{e1}. This gives
\be\label{pf3}
\int d\eta\,a^4 \fr{dV(\phi,\a)}{d\a}=\fr{1}{8\pi^3}\int d\eta d^3k \fr{d R_{\mu}}{d\a}\fr{|\chi_k(\eta)|^2}{Z(\phi,\a)},
\ee
which trivially replaces \eq{pf}. In this new scheme the optimized cutoff function becomes
\be\label{rr}
R_\a(k^2)=Z(\phi,\a)(\a^2a^2-k^2)\th(\a^2a^2-k^2)
\ee
and using \eq{rr} in \eq{me} gives the new mode equation. Unlike \eq{me2}, the mode equation becomes piecewise defined in time for given constants $\a$ and $k$:
\bea
&&a>\fr{k}{\a}:\hs{3}\chi''_\a+\left[\a^2a^2-\fr{a''}{a}+\fr{a^2}{2Z(\phi,\a)}\fr{d^2V}{d\phi^2}\right]\chi_\a=0,\nn\\
&&a<\fr{k}{\a}:\hs{3}\chi''_\a+\left[k^2-\fr{a''}{a}+\fr{a^2}{2Z(\phi,\a)}\fr{d^2V}{d\phi^2}\right]\chi_\a=0.\label{dm}
\eea
This equation must be solved in two different regions and the pieces should be matched at $a=k/\a$. The mode function must be continuos and should have a continuos first time derivative. Even though the equation in the range $a>k/\a$ is free from $k$, the matching generates $k$ dependence. As a result it is not possible to carry out the momentum integral in \eq{pf3} without solving for $\chi_\a$ explicitly. Note that the anomalous dimension $\eta_\phi$ is the same for the flow parameters $\mu$ and $\a$. 

One may think that the flow equations with comoving and physical cutoffs should agree with each other  since they look like the reparametrization of the same curve. This is not correct because they correspond to coarse graining of different scales in time and the system does not have time translation invariance. The exact quantum effective actions must be the same when the cutoffs are removed but the flows might differ when an approximation is employed. In flat space, the comoving cutoff is analogous to a time-dependent scheme which eliminates different physical scales at different times. In the next section, we analyze both of the flow equations in detail. As we will see, while the comoving cutoff scheme faces the trans-Plankian problem, the physical cutoff scheme has a peculiar IR behavior. 

\section{ERG Flow equations during Inflation} \label{3}

In this section, we analyze \eq{pf2} and \eq{pf3} for a scalar in an inflationary universe and take
\be
a=-\fr{1}{H\eta},
\ee
where $H$ is the Hubble constant. The inflation is assumed to last for $N$ e-folds and typically $N\geq 60$. We apply the LPA,\footnote{In flat space, the LPA amounts to take the potential independent of four-momentum and to neglect the anomalous dimension. Since we are not  Fourier transforming the time direction, in our case the LPA is equivalent to $\eta$-independence of the potential.} namely we take $V(\phi,\mu)$ to be independent of $\eta$ (recall it may have an explicit time dependence) and ignore the possible contributions of anomalous dimension by setting 
\be
Z(\phi,\mu)=1,\hs{5}\eta_\phi=0,
\ee
which leads to a closed differential equation system involving only the parameters of the potential. The LPA is known to be a good approximation for many applications in flat space \cite{rev1}. 

We start with the comoving cutoff scheme. In the LPA,  \eq{me2} becomes the free-field mode equation with a modified mass term and the comoving momentum value. Therefore we take
\be
\chi_\m=\chi_k^{BD}(k\to\mu),
\ee
where $\chi_k^{BD}$ is the mode function in the Bunch-Davies vacuum. This gives
\be\label{mm} 
\chi_\mu=\sqrt{\fr{-\pi\eta}{4}}\,e^{i\fr{\pi}{2}(n+\fr12)}\,H^{(1)}_n(-\mu\eta).
\ee
where 
\be\label{n}
n=\sqrt{\fr94-\fr{1}{2H^2}\fr{d^2V}{d\phi^2}}
\ee
and $H_n^{(1)}$ is the first Hankel function. Without loss of any generality one may take $\eta_i=-1/H$ (so that $a(\eta_i)=1$) and write 
\be
\eta_f=\fr{-1}{H}e^{-N},
\ee
where, as noted above, $N$ denotes the number of e-folds. The complex conjugate of $\chi_\m$ can be found as 
\be\label{mmc}
\chi_\m^*=\sqrt{\fr{-\pi\eta}{4}}\,e^{-i\fr{\pi}{2}(n+\fr12)}\,H^{(2)}_n(-\mu\eta),
\ee
which is valid both for real and complex $n$. 

\begin{figure}
\centerline{\includegraphics[width=8cm]{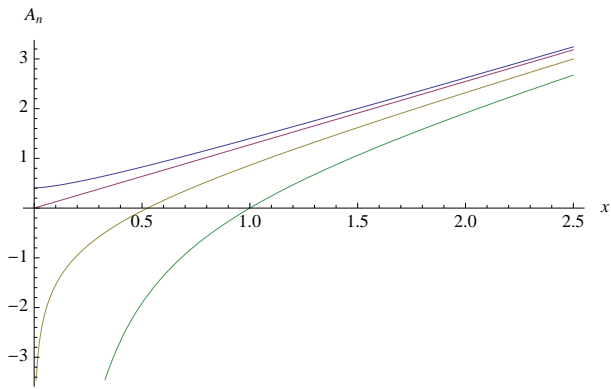}}
\caption{The graph of $A_n(x)$ for $n=0,1/2,1,3/2$, from top to bottom, respectively. For $n\leq 1$, the function diverges as $x\to0$.} 
\label{fig1}
\end{figure}

Using \eq{mm} in \eq{pf2} and  carrying out $\eta$-integrals yield
\be
\fr{dV}{d\mu}=\fr{H\mu^2}{8\pi(e^{3N}-1)}\left[A_n(\mu/H)-A_n(\mu e^{-N}/H)\right],
\ee
where the function $A_n$ is given by 
\bea
A_n(x)=&&x^2\left[J_n(x)^2+Y_n(x)^2\right.\nn\\
&&\left.-J_{n-1}(x)J_{n+1}(x)-Y_{n-1}(x)Y_{n+1}(x)\right].
\eea
Here $J$ and $Y$ are the Bessel functions of first and second kinds, respectively. $A_n(x)$ is an increasing function of its argument for real $n$, with a somehow weak dependence on the index $n$ (see Fig. \ref{fig1}). For later use let us note the following asymptotic behavior: 
\be\label{a1}
A_n(x)=\fr{4x}{\pi}+\fr{1-4n^2}{2\pi x}-\fr{9-40n^2+16n^4}{32\pi x^3}+{\cal O}\left(\fr{1}{x^4}\right)
\ee
as $x\to\infty$, and 
\bea
&&A_n(x)=\left(\fr{4n\cot(n\pi)}{\pi}+{\cal O}(x)\right)\label{a2}\\
&&+x^{-2n}\left(\fr{4^n[\C(n)^2-\C(n-1)\C(n+1)]}{\pi^2}x^2+{\cal O}(x^3)\right)\nn
\eea
as $x\to0$ for $\textrm{Re}(n)>0$. 

Defining a new dimensionless parameter as 
\be\label{k}
\k\equiv e^{-N}\fr{\mu}{H},
\ee
and assuming $e^N\gg 1$, the flow equation becomes
\be\label{s}
\fr{dV}{d\k}\simeq\fr{H^4}{8\pi}\k^2\left[A_n(e^{N}\k)-A_n(\k)\right],
\ee
which is the beta functional of the potential, i.e. one may Taylor expand both sides of \eq{s} in $\phi$ to read the beta functions. Note that  the right hand side of \eq{s} depends on the potential through the index $n$ given in \eq{n}. 

Before analyzing the solutions of \eq{s} for a quartic potential, we would like to comment on a few important points. Recall that $\mu$ has been introduced as a comoving scale, $\eta_i$ is chosen such that $a(\eta_i)=1$ and $a(\eta_f)=e^N$. Thus, $\mu$ corresponds to the physical scale in the beginning of inflation 
and $\k$ is the physical scale at the end of inflation. Since physics after inflation is our main concern, $\k$ is the {\it scale of interest} and UV/IR limits must refer to the variable $\k$. 

In a realistic scenario $N\geq 60$ or so, and \eq{s} clearly exhibits the trans-Plankian problem \cite{tr1,tr2,tr3}. Namely,  when $\k$ is near Planck scale, there appears another scale in the problem $e^N\k$ which is trans-Plankian. On the other hand, for $\k$ near IR, $e^N\k$ can be UV and thus \eq{s} also shows UV/IR mixing, which is a peculiar property of de Sitter physics \cite{pol2,pol1}.

As noted above $A_n(x)$ is an increasing function of its argument. From the basic construction of the ERG flow, one must imagine starting the flow at some UV scale and ending up at an IR scale (in principle at $\k=0$). We therefore see from \eq{s} that $V$ decreases\footnote{This can already be observed in \eq{pf2} for $\eta_\phi<5$.}  along the flow for any given value of $\phi$.

It is crucial to emphasize that the quantum action is defined once and for all in the interval $(\eta_i,\eta_f)$ and the action itself does not evolve in time. Therefore, in the above construction the ERG flow should not be interpreted as an evolution in time. However, due to the external time dependence coming from the background,  the parameters in the effective action are expected to vary with time. 

It is interesting to consider  an {\it almost massless} field so that $m(\k)^2/H^2\equiv d^2V/d\phi^2|_{\phi=0}/H^2\sim0$, which actually requires an extreme fine tuning of the parameters. In that case, one may set $n=3/2$ and integrate \eq{s} for $\phi=0$ to obtain
\be\label{l0}
\L(\k)\simeq\fr{H^4}{8\pi^2}\left[e^N(\k^4-\k_0^4)-2(\k^2-\k_0^2)\right]+\L_0,
\ee
where $\L\equiv V|_{\phi=0}$ and $\L_0$ is an integration constant chosen such that $\L(\k_0)=\L_0$. Thus  the ERG flow gives
\be\label{l1}
\L(0)\simeq\L_0-\fr{e^N}{8\pi^2}H^4\k_0^4,
\ee
which shows that the the bare cosmological constant $\L_0$ is screened by a huge negative amount.

The above result shows how ERG might be used to address some crucial details in the cosmological constant problem. Imagine that one calculates quantum corrections to the cosmological constant coming from a field theory in an expanding universe. In such a calculation the easiest way to regulate infinities is to introduce a cutoff, which is somehow ambiguous since the scales redshift  with the expansion. In our case, for example, inflation generates a huge hierarchy of scales in time and the physics may  depend sensitively on how the cutoff is chosen. The ERG equation \eq{s} solves this ambiguity, since it actually gives a cumulative effect overtime and as a result $e^N$ factor in \eq{l1} arises. 

Let us now truncate the potential as
\be\label{tr}
V=\L+m^2\phi^2+\l\phi^4.
\ee
To get the beta functions for $\L$, $m^2$ and $\l$, one must expand \eq{s} in powers of $\phi$ and equate the coefficients. In flat space studies, the cosmological constant term $\L$ is not considered in the ERG flows. In the hierarchy of parameters, $\L$ is the lowest order member, i.e. its flow is affected by other parameters but it does not affect  the flow of any parameter. Therefore, ignoring it in flat space causes no problem other than the fact that the bare cosmological constant must be fine tuned to neglect any backreaction effects. Here, aiming to apply the ERG techniques to the cosmological constant problem, we keep $\L$ to see its scale dependent running. Since we ignore backreaction effects, an implicit fine tuning is also assumed so that $\L$ becomes small compared to the background cosmological constant. 

\subsubsection*{UV Regime}

It is difficult  to analyze the beta functions arising from the flow equation \eq{s} exactly, mainly because of flow varying indices of the Bessel functions. However, it is possible to simplify the equations in certain limits, which helps to understand the qualitative flow behavior. In the UV regime for $\k\gg1$ (recall that we have already assumed $e^N\gg1$) using the asymptotic formula given in \eq{a1} one finds
\bea
&&\k\gg1:\nn\\
&&\k\del_\k\L\simeq\fr{H^4e^N}{2\pi^2}\k^4-\fr{H^2}{4\pi^2}m^2\k^2+\fr{1}{16\pi^2}m^4,\nn\\
&&\k\del_\k m^2\simeq-\fr{3H^2}{2\pi^2}\l \k^2+\fr{3}{4\pi^2}m^2\l,\label{fl2}\\
&&\k\del_\k\l\simeq\fr{9}{4\pi^2}\l^2.\nn
\eea
Let us note that in obtaining \eq{fl2} no condition on $m^2$ is imposed, i.e. it can be any real number including the negatives. From \eq{k}, the $H$ dependence in the above equations becomes redundant. The flow equations for $m^2$ and $\l$  are similar to the flat space counterparts and the only imprint of the expansion of the universe is the $e^N$ term that appears in the beta function of $\L$.  

The last equation in \eq{fl2} can be solved as
\be\label{tri}
\l(\k)\simeq\fr{\l_0}{1-\fr{9\l_0}{4\pi^2}\ln(\k/\k_0)},
\ee
which is identical to the flat space behavior. The usual ``triviality" argument follows from the fact that \eq{tri} implies $\l(\k)\to0$ as $\k\to0$. However, in our case \eq{tri} is valid for $\k\gg1$ and as we will see shortly the flow becomes quite different at IR. 

Since $\l$ runs logarithmically for $\kappa\gg1$, one may take $\l\simeq\l_0$ in the second equation in \eq{fl2} to get
\be\label{mf}
m^2(\k)\simeq-\fr{3\l_0}{4\pi^2}H^2\k^2+\left(\fr{\k}{\k_0}\right)^{3\l_0/4\pi^2}\left[m_0^2+\fr{3\l_0}{4\pi^2}H^2\k_0^2\right],
\ee
where $m^2(\k_0)=m_0^2$. This is again consistent with the flat space behavior, i.e. the renormalized mass quadratically depends on the UV scale. Note that as one lowers $\k$, $m^2$ gets larger. 

From \eq{mf}, one finds that $m^2\sim {\cal O}(\k^2)$ and using this in the first equation in \eq{fl2}, the first term in the right hand side can be seen to give the largest contribution. Integration then yields
\be
\L(\k)\simeq\fr{H^4}{8\pi^2}\left[e^N(\k^4-\k_0^4)\right]+\L_0,
\ee
which is consistent with \eq{l0} for  $\k\gg1$. Therefore, in the UV regime the greatest contribution to the cosmological constant is independent of the bare parameters in the potential, it is in some sense purely {\it geometrical} and arises due to the expansion of the universe. 

\subsubsection*{IR regime for a light scalar}

Consider now the next interval in the flow where one has $1\gg\k\gg e^{-N}$. When $N$ is large, this interval  includes even deep IR excitations. For generic values of $m^2$, the expansion of the function $A_n$ in \eq{s} involves the derivatives of the Bessel functions with respect to their arguments and the expressions become very complicated. However, the formulas simplify considerably for a light scalar, which has $|m^2(\k)|/H^2\ll1$ since the index $n$  can be expanded around $n=3/2$. 
 
In that case, using \eq{a2} one finds  
\be\label{l2}
\k\del_\k\l\simeq\fr{\k^2}{6\pi^2}\left[a_1+a_2\ln(\k)+a_3\ln(\k)^2\right]\l^2,
\ee
where the constants involve factors like the Euler's number coming from the expansion of the Bessel functions and they are given by $a_1\simeq40$, $a_2\simeq21$ and $a_3=24$. The flow of $\l$ changes drastically compared to \eq{fl2}, and \eq{l2} shows that 
\be
\l\to\textrm{const.}\hs{4}\textrm{as} \hs{4}\k\to0
\ee
invalidating the triviality argument. Indeed, the variation of  $\l$ is small since $\k\ll1$ in \eq{a2} 
and one may take
\be\label{tr2}
\l(\k)\sim\textrm{const.}\sim\fr{\l_0}{1+\fr{9\l_0}{4\pi^2}\ln(\k_0)}:\hs{4}1\gg\k\gg e^{-N},
\ee
where $\l$ is matched with \eq{tri} at $\k=1$.  

In the same interval, the ERG flow equations for $m^2$ and $\L$ become
\bea
\k\del_\k m^2\simeq&&H^2\k^2[b_1+b_2\ln(\k)]\l\nn\\
&&+\k^2[b_3+b_4\ln(\k)+b_5\ln(\k)^2]m^2\l\label{m2}
\eea
and 
\bea
\k\del_\k\L\simeq&&\fr{H^4}{2\pi^2}\left[e^N\k^4+\k^2\right]+H^2\k^2[c_1+c_2\ln(\k)]m^2\nn\\
&&+\k^2[c_3+c_4\ln(\k)+c_5\ln(\k)^2]m^4,\label{L2}
\eea
where $b_i$ and $c_i$ are numbers of order unity (note the expansion in powers of $m^2/H^2$). In the beta function of $\L$, we see that the first term containing $e^N$ initially gives the largest contribution, but as $\k$ is lowered $\k^2$ terms become dominant. Since the right-hand sides of \eq{m2} and \eq{L2} are small in Hubble units, there appears no {\it significant contribution} to $m^2$ and $\L$ in this IR region with $1\gg\k\gg e^{-N}$.

Finally one may consider the extreme IR regime with $\k \ll e^{-N}$.  Using the asymptotic formula \eq{a2} one sees that the beta functions  for $m^2$ and $\l$ become identical to \eq{l2} and \eq{m2}, where all the terms in the right hand sides are multiplied by $-1$. Similarly for $\L$, one gets \eq{L2} with no $e^N$ term and all others in the right hand side change sign. These can be seen by observing that while in \eq{s} $A_n(e^N\k)$ is much larger than $A_n(\k)$ for $1\gg\k\gg e^{-N}$, the opposite is true for $\k \ll e^{-N}$. As a result, the flow becomes trivial and the runnings are  irrelevant in the deep IR regime with $\k \ll e^{-N}$.

\subsection*{Physical Cutoff}

To analyze the beta functional in the physical cutoff,  one must first solve \eq{dm}. For convenience, we scale $\a$ as
\be\label{ah} 
\a\to H\a
\ee
so that it becomes dimensionless. The solution of \eq{dm} can be found as  
\be
\chi_\a=
\begin{cases}\sqrt{\fr{-\pi\eta}{4}}\,e^{i\fr{\pi}{2}(n+\fr12)}\,H^{(1)}_n(-k\eta):\hs{1}\eta<-\fr{\a}{k}\\
c_1(-k\eta )^{1/2-r}+c_2(-k\eta)^{1/2+r}:\hs{1}\eta>-\fr{\a}{k}\end{cases}\label{ps}
\ee
where $n$ is given by \eq{n} and 
\be\label{rphi}
r=\sqrt{\fr94-\a^2-\fr{1}{2H^2}\fr{d^2V}{d\phi^2}}.
\ee
The normalization of the solution for $\eta<-\a/k$ is fixed by referring to the Bunch-Davies vacuum. Demanding that the mode function and its first derivative to be continuous at $\eta=-\a/k$ fixes the integration constants as 
\bea
&&c_1=\sqrt{\fr{\pi}{16k}}e^{i\fr{\pi}{2}(n+\fr12)}\a^r\left[-\fr{\a}{r}H_n^{'(1)}(\a)+H_n^{(1)}(\a)\right]\\
&&c_2=\sqrt{\fr{\pi}{16k}}e^{i\fr{\pi}{2}(n+\fr12)}\a^{-r}\left[\fr{\a}{r}H_n^{'(1)}(\a)+H_n^{(1)}(\a)\right]
\eea
where prime denotes derivative with respect to the argument. 

Using \eq{ps} in \eq{pf3} and carrying out the momentum integral, one sees that the $\eta$ integral in the right-hand side of \eq{pf3} becomes $\int d\eta a^4$, which is exactly the same with the left hand side. As a result, the time integrals on both sides cancel out each other and \eq{pf3} becomes independent of $\eta_i$, $\eta_f$, and hence the number of e-folds $N$. The ERG equation with the physical cutoff does not suffer from the large hierarchy of scales during inflation. The modes are eliminated gradually in time and this solves the trans-Plankian problem. 

\begin{figure}
\centerline{\includegraphics[width=8cm]{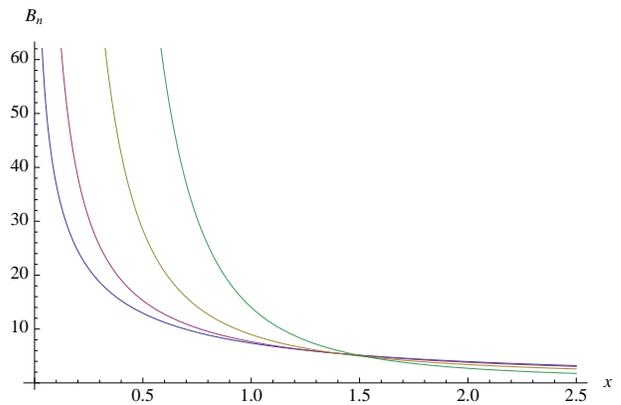}}
\caption{The graph of $B_n(x)$ for $n=0,1/2,1,3/2$, from left to right, respectively.} 
\label{fig2}
\end{figure}

After some algebra we find that \eq{pf3} becomes
\be\label{bfp}
\fr{dV}{d\a}=\fr{H^4}{48\pi}\a^4\left(\a^2+\fr{1}{2H^2}\fr{d^2V}{d\phi^2}\right)^{-1}\,B_n(\a),
\ee
where
\bea
&&B_n(\a)=2\a^2H_{n-1}^{(1)}(\a)H_{n-1}^{(2)}(\a)\\
&&-\a(3+3n)\left(H_{n-1}^{(1)}(\a)H_n^{(2)}(\a)+H_n^{(1)}(\a)H_{n-1}^{(2)}(\a)\right)\nn\\
&&+(9+6n-2\a^2)H_n^{(1)}(\a)H_n^{(2)}(\a).\nn
\eea
Eq. \eq{bfp} is the beta functional of the potential in the physical cutoff scheme. Here, $B_n(x)$ is a monotonically decreasing function of its argument for real $n$ (see Fig. \ref{fig2}), which has the following asymptotic expansions: 
\bea
B_n(x)=&&\fr{26-8n^2}{\pi x}-\fr{27+16n^2(n^2-7)}{4\pi x^3}\label{ba1}\\
&&+\fr{3(441-1996n^2+944n^4-64n^6)}{64\pi x^5}+{\cal O}\left(\fr{1}{x^6}\right),\nn
\eea
as $x\to\infty$ and 
\bea
B_n(x)=&&\left(\fr{2[8n^2-9]\cot(n\pi)}{n\pi}+{\cal O}(x)\right)\label{ba2}\\
&&+x^{-2n}\left(\fr{3[3+2n]4^n\C[n]^2}{\pi^2}+{\cal O}(x^2)\right),\nn
\eea
as $x\to0$ for $\textrm{Re}(n)>0$. 

Even though the trans-Plankian problem is solved, \eq{bfp} may suffer from peculiar IR behavior. If $\a^2+m^2(\a)$ vanishes\footnote{Recall that one should  expand both sides of \eq{bfp} around $\phi=0$ and $m^2\equiv d^2V/d\phi^2|_{\phi=0}/2$.} at some point along the flow, then \eq{bfp} may become singular depending on the behavior of the function $\a^2+m^2(\a)$ around its zero. For instance, as $\a\to0$ one finds that 
\be\label{ld}
 \left.\fr{dV}{d\a}\right|_{\phi=0}\sim C\fr{\a^{4-2\sqrt{9/4-m^2/H^2}}}{\a^2+m^2/H^2}+...
 \ee
where $C$ is some constant. If $m^2$ vanishes faster than or equal to $\a^2$ as $\a\to0$, i.e. $m^2\sim \a^p$ with $p\geq 2$, then \eq{ld} becomes logarithmically divergent at $\a=0$. This is related to the well-known IR divergence of a massless scalar in de Sitter space, which shows up as a singularity in the RG flow. 

It is again interesting to consider an almost massless field before discussing the generic case below. After neglecting mass and setting $\phi=0$, which gives $n=3/2$, \eq{bfp} can be integrated to yield
\bea
\L(\a)=&&\fr{H^4}{24\pi^2}\left[(\a^4-\a_0^4)+4(\a^2-\a_0^2)+18\ln(\a/\a_0+\e)\right]\nn\\
&&+\L_0, \label{l30}
\eea
where $\L=V|_{\phi=0}$, $\L_0$ is an integration constant and $\e\sim m^2/H^2\sim0$. Thus the ERG flow gives
\be\label{l3}
\L(0)\simeq\L_0-\fr{H^4}{24\pi^2}\left[\a_0^4+4\a_0^2-18\ln(\e)\right].
\ee
As in \eq{l1}, we see a large screening of the cosmological constant, of the order of $H^4\a_0^4$, where $\a_0\gg1$. However, contrary to \eq{l1} the screening becomes independent of the number of e-folds, i.e. it can be thought to arise as a result of formulating the theory in the eternal de Sitter space. For a massless field, $\e=0$ and the flow diverges logarithmically, which is the IR problem mentioned above. 

\subsubsection*{UV regime}

Once more the beta functional \eq{bfp} is complicated to solve exactly and as before we first analyze it in the UV regime. Truncating the potential as in \eq{tr} and using the expansion formula \eq{ba1}, for 
$\a\gg1$ (and for any real value of $m^2$) \eq{bfp} gives
\bea
&&\a\gg1:\nn\\
&&\a\del_\a\L\simeq\fr{H^4}{6\pi^2}\a^4-\fr{H^2}{12\pi^2}m^2\a^2+\fr{1}{16\pi^2}m^4,\nn\\
&&\a\del_\a m^2\simeq-\fr{H^2}{2\pi^2}\l \a^2+\fr{3}{4\pi^2}m^2\l,\label{flp}\\
&&\a\del_\a\l\simeq\fr{9}{4\pi^2}\l^2.\nn
\eea
After a constant rescaling $\a\to\sqrt{3}\a$, \eq{flp} becomes identical to \eq{fl2} except the first term in the beta function of $\L$. Thus, in the UV regime the comoving and the physical cutoff schemes agree on the runnings of $m^2$ and $\l$. They also agree on $m^2$ and $\l$ dependent contributions to $\L$-flow but they disagree\footnote{We checked very carefully that the first terms in the beta functions of $\L$ in \eq{fl2} and \eq{flp} are  different.} on the ``geometrical" term. For $\a\gg1$ the solutions are given by \eq{tri} and \eq{mf}, where $\k$ is replaced by $\sqrt{3}\a$, therefore as one lowers the cutoff, $\l$ decreases and $m^2$ increases for arbitrary initial data. 

\subsubsection*{IR regime for a light scalar}

In the IR regime for $\a\ll1$, we again focus on a light scalar obeying $|m^2|/H^2\ll1$, so that the index $n$ can be expanded around $n=3/2$. Using the asymptotic formula \eq{ba2}  we find  from \eq{bfp} that 
\be\label{plf}
\a\del_\a\l\simeq\fr{1}{6\pi^2}\fr{\a^2\l^2}{(\a^2+m^2/H^2)^3}\left[162+...\right]
\ee
where the dots denote the terms like $\a^2$, $\a^4$, $\a^2\ln(\a)$ etc. that vanish as $\a\to0$. Contrary to the  previous flow given in \eq{l2}, the IR behavior of $\l$ depends sensitively on the running of $m^2$. This is because of $(\a^2+d^2V/(2H^2d\phi^2))^{-1}$ term in \eq{bfp} which does not contain any additive constants. For comparison note that in \eq{s}, $m^2$ only appears through the index $n$ and consequently it is always combined with the Hubble's constant $H$ in the form $9/4-m^2/H^2$.  Therefore, \eq{plf} must be examined with the running of $m^2$, which can be determined from \eq{bfp} as 
\bea
&&\a\del_\a m^2\simeq\fr{H^2}{36\pi^2}\fr{\a^2\l}{(\a^2+m^2/H^2)^2}\times\label{pmf}\\
&&\left[-162+\fr{m^2}{H^2}\left(\fr{m^2}{H^2}+2\a^2\right)(d_1+d_2\ln(\a)+d_3\ln(\a)^2\right],\nn
\eea
where $d_1\simeq34$, $d_2\simeq-56$ and  $d_3=36$.  

Examining \eq{plf} and \eq{pmf}, one may see that there is a solution for which $\l$ and $m^2$ approach nonzero constants:    
\be\label{pfs1}
\l\to\l_R,\hs{4}m^2\to m^2_R\hs{4}\textrm{as} \hs{4}\a\to0.
\ee
Such a solution exists since the right-hand sides of the differential equations \eq{plf} and \eq{pmf} vanish as $\a\to0$ if $m^2$ approaches a nonzero constant. The measurable parameters $m^2_R$ and $\l_R$ are determined by the initial conditions that must be supplied for the flow. This IR behavior is consistent with the one obtained in the comoving cutoff scheme. 

However, there is another interesting solution of \eq{plf} and \eq{pmf}, which has completely different IR characteristics.  Assuming that
\be
m^2\simeq m_0^2\a^2,
\ee 
\eq{plf} gives
\be
\l\simeq\fr{4\pi^2}{27}\left(1+\fr{m_0^2}{H^2}\right)^3\a^4.
\ee
Using this in \eq{pmf} fixes the value of $m_0^2$ as
\be
m_0^2=-\fr{1}{4}H^2.
\ee
Therefore \eq{plf} and \eq{pmf} admit an IR {\it scaling} solution for $\a\ll1$, which is approximately given by
\be\label{pfs2}
\l\simeq \fr{\pi^2}{16}\a^4,\hs{4}m^2\simeq-\fr14 H^2\a^2\hs{4}\textrm{as} \hs{4}\a\to0.
\ee
Note that in deriving \eq{plf} and \eq{pmf} we have assumed $|m^2|/H^2\ll1$ and the scaling solution satisfies this requirement. 

On the other hand, the beta function of the cosmological constant $\L$ can be determined from \eq{bfp} as
\bea
&&\a\del_\a\L\simeq \fr{H^4}{216\pi^2}\fr{\a^2}{\a^2+m^2}\left[162+\label{pccf}\right.\\
&&\left.\fr{m^2}{H^2}[e_1+e_2\ln(\a)]+\fr{m^4}{H^4}[e_3+e_4\ln(\a)+e_5\ln(\a)^2]\right]\nn,
\eea
where $e_1\simeq-97$, $e_2\simeq108$, $e_3\simeq84$, $e_4\simeq-102$ and $e_5=36$. For the first solution \eq{pfs1}, the ERG flow \eq{pccf} can be integrated to yield
\be
\L\simeq\fr{H^4}{216\pi^2}\fr{\a^2}{m_R^2/H^2} +\L_0.
\ee
As $\a\to0$, the change in $\L$ is negligible and the flow becomes irrelevant at IR. However,  the integration of \eq{pccf} for the scaling solution \eq{pfs2} gives
\be
\L\simeq\fr{H^4}{162\pi^2}\ln(\a)+\L_0,
\ee
which shows that $\L$  receives a large (divergent) negative IR contribution. Since $\L$ diverges as $\a\to0$, one may prefer to keep it small but finite, which implies a physically observable nonzero value for $m^2$. If $m_{obs}^2$ denotes this measurable value, the scaling solution \eq{pfs2} gives a minimum for $\a$ as
\be
\a_*= 2\sqrt{-\fr{m_{obs}^2}{H^2}}.
\ee
The change in $\L$ in the IR regime is given by
\be
\D\L\simeq\fr{H^4}{162\pi^2}\ln(\a_*),
\ee
which should be added to the amount of UV screening. Note that when $m^2<0$ the potential minima becomes degenerate at $\phi^\pm_{min}=\pm\sqrt{-m^2/(2\l)}$ and there emerges a global $Z_2$ symmetry, which is broken by the nonzero scalar vacuum expectation value.  

\section{Discussion}

In this section, we would like to elaborate on the interpretation and possible implications of our findings. As noted previously, to obtain {\it the full quantum effective action} the cutoff must be sent to zero, and thus the comoving and the physical cutoff schemes should agree with each other in the exact quantum picture. However, the action is usually truncated as we did above and consequently one is not dealing with the full quantum action but instead an {\it effective description} relevant to a given physical situation. In that case, the ERG flow parameter must be set to {\it the scale of interest} in the problem. 

It is important to emphasize that we take a {\it fixed} gravitational background and neglect any possible backreaction effects.  The background geometry sets a scale, which is the Hubble constant $H$,  and in fixing the value of the ERG flow parameter, this scale must be taken into account. 

In the light of the above comments, let us, for example, elucidate \eq{l0} and \eq{l1}, which give the ERG flow of a massless scalar in an inflationary spacetime. First, recall that although $\k$ is defined to give the physical scale at the end of inflation, it is actually a {\it comoving} scale measured in units of $H$. Therefore \eq{l0} should be thought to give the effective value of $\L$ at the given comoving scale $\k$, provided that $\L_0$ is the effective vacuum energy density at the comoving (UV) scale $\k_0$. This standard interpretation becomes subtle in the comoving cutoff scheme due to the time dependence of the background, which shows up as $N$-dependence in \eq{l0}. Thus, the RG flow fixes the effective value of $\L$ at a given comoving scale at a given time and \eq{l0} is expressed to yield the scale dependence of $\L$ at the end of inflation. Specifically, \eq{l1} gives the total screening of $\L$ from UV to IR regimes at the end of inflation (since $\k$ is defined to give the physical scale at the end of inflation) and if, for instance, one would like to determine the impact of the vacuum energy density at cosmological scales, it is $\L$ and not $\L_0$ that must be taken into account. Similarly, if one is interested in the UV physics at scale $\k_0$ at the end of inflation, the value of $\L_0$ must be used as the corresponding vacuum energy density. 

The RG flow \eq{l30} in the physical cutoff scheme is much more easy to interpret. As before, \eq{l30} gives the effective value of $\L$ at the physical scale $\a$ but this time $\L(\a)$ turns out to be time independent. Here, the time dependence is absorbed in the flow parameter, which is chosen to be the physical scale redshifting with expansion. One may imagine that  the effective vacuum energy density of the quantum scalar decreases as it flows from sub-Hubble UV scale to the super-Hubble IR scale due to the constant spacetime curvature acting as a potential barrier. 

The above results  describe how the effective vacuum energy density of a test scalar changes from UV to IR in a {\it fixed} inflationary background and it is encouraging to see that the RG flow implies a fall off, i.e. the classical gravitational field screens the vacuum energy density at large scales during inflation. However, at this point it is not possible to claim that this provides a realistic screening to give a graceful exit mechanism for inflation as in, e.g., \cite{exit}. To verify this,  backreaction effects must be taken into account, i.e. the gravitational dynamics should be determined in the presence of the matter/scalar (and possibly metric) fluctuations (alternatively, one may also include the gravitational degrees of freedom in the ERG flow analysis to obtain an effective action including the gravitational degrees of freedom  as in \cite{ergc1,ergc2}). 

Keeping these in mind, one may try to understand possible implications of our findings for the cosmological constant problem. In our analysis, the RG scale is considered to be a free parameter which can be set according to the physical problem at hand. In determining the cosmological impact of $\L$, it is natural to set the {\it physical} RG scale at the instantaneous Hubble expansion rate as in \cite{exit,c11} (see also \cite{c12}, which uses the scale factor of the universe to fix the RG scale). This motivates the physical cutoff scheme and let us consider the flow \eq{l30}.  In writing \eq{l30}, we had scaled the flow parameter $\a$ in \eq{ah} to make it a dimensionless parameter.  To avoid any complication, one may  simply rescale $\a\to \a/H$ to obtain
\bea
\L(\a)=&&\fr{1}{24\pi^2}\left[(\a^4-\a_0^4)+4H^2(\a^2-\a_0^2)\right.\nn\\ && \left.+18\ln(\a/\a_0+\e)\right]+\L_0.
\eea
Note that this rescaling  corresponds to a simple reparametrization of the same RG trajectory that does not change physics. Let us now assume that this equation is still valid when $H$ is changing (at least slowly) and the cosmological impact of the vacuum energy is effective at the Hubble scale so that $\a\simeq H$. Assuming further that $\a\ll\a_0$, the effective cosmological constant at the Hubble radius becomes
\be\label{d1}
\L(H)\simeq \L_0-
\fr{\a_0^4}{24\pi^2}-\fr{H^2\a_0^2}{6\pi^2},
\ee
where we ignore the logarithmic term to keep the discussion simple. The UV scale can naturally be identified with the Planck scale, i.e. $\L_0\simeq M_P^4$ and $\a_0\simeq M_P$. The first subtracted term in \eq{d1} is the flat space contribution and the second one is related to the expansion of the universe. As usual, the characteristic quartic quantum contribution to the cosmological constant appears with the opposite sign in the RG flow analysis, but the sign of the last geometric term is nontrivial.  In principle, the bare cosmological constant $\L_0$ can (mostly) be canceled out by the $\a_0^4$ term by  fine tuning. On the other hand, \eq{d1} shows that for large $H$ the screening of $\L$ is large and as $H$ decreases $\L$ increases. Since $\L(H)$ is supposed to act like a cosmological constant in Einstein's equations, \eq{d1} implies the existence of an ``equilibrium" value for $H$ (if the last term in \eq{d1} would appear with a positive sign, then this would indicate an instability, i.e growing $H$ would give an increasing $\L(H)$, which would then force $H$ to grow more via Einstein's equations). Although the RG flow indicates a screening of the vacuum energy density for a massless scalar reducing the value of the bare cosmological constant, extra ingredients are necessary to solve the graceful exit problem.

One possible modification is to consider a massive field, which is supposed to yield $m^4$ and $m^2H^2$ terms in the ERG equations. Indeed, these terms can already be seen to arise in the flow \eq{L2}, however they are not expected to give significant contributions since we have assumed $m\ll H$. On the other hand, our results show that the ERG analysis can provide a theoretical basis for the phenomenological models studied in \cite{ph1,ph2,ph3,ph4}, which mainly consider cosmological consequences of a vacuum energy density fixed by a polynomial of the Hubble parameter. 

For small $H$, the screening of the bare cosmological constant is small by \eq{d1}. This is somehow expected since our analysis uncovers the impact  of the spacetime curvature on the ERG flow. Therefore, at least for a massless field, our results are not useful to understand the accelerated expansion of the universe today (to avoid any confusion let us note that the extreme IR regime discussed in section \ref{3} does not refer to our universe today, but it arises in the inflationary regime due to the exponential hierarchy of scales related to the trans-Plankian problem). However, it would be interesting to consider the derivative expansion in an expanding universe along the lines of \cite{mor} to determine the form of the effective potential near a {\it fixed point}, which would give a nontrivial dependence on the Hubble parameter.   

\section{Conclusions} 

In this paper, we obtain the ERG flow equations for the in-out and in-in quantum effective actions corresponding to  a self-interacting real scalar field in an expanding cosmological spacetime, which are given by \eq{erg1} and \eq{erg2}, respectively. As in flat space, the ERG equations can be derived using the path integral representation, however, formulating the theory in a cosmological background gives rise to some key differences. Although it is a minor technical detail, the derivation must be carried out in Lorentz signature since it is not known how to Wick rotate a general cosmological spacetime. Moreover, one can only Fourier transform the three spatial dimensions. The ERG equations involve the exact propagator that should be obtained from the quantum effective action by \eq{g1} and \eq{g2}. In flat space, at least in some approximations, this is a trivial procedure where one simply takes the inverse of a polynomial in momentum space. In the expanding universe, the Green function must be constructed by referring to a vacuum state. 

May be the most striking difference compared to the flat space ERG calculations is the need for an extra input for the interpretation of the quantum effective action. One usually thinks of an action to be defined for the whole spacetime. Moreover the fields must be imposed to obey certain boundary/fall off  conditions for the variational principle to work. In the absence of nice asymptotic regions as in a generic cosmological spacetime, defining a well behaved action and a variational principle can be problematic. To avoid technical complications, we follow \cite{pol1} and assume that the cosmological background is only active for a finite time interval $(\eta_i,\eta_f)$, which is glued to suitable past and future asymptotic regions. Supposing further that the interactions are turned off outside this interval, the action functionals, including the quantum effective action, can be thought to de defined in the region $(\eta_i,\eta_f)$. Such a description is suitable for inflation, where $\eta_i$ and $\eta_f$ mark the beginning and the ending of the exponential expansion. In that case, the quantum effective action encodes the imprints of the classical gravitational field on the quantum dynamics of the scalar. 

Yet another difference with the flat space ERG analysis is the existence of comoving and physical cutoff schemes. In ERG theory, the term scheme usually refers to a choice of a cutoff function. In an expanding universe one can consider two main coarse graining procedures with respect to the comoving or the physical scales. Although, full quantum effective actions must agree as the cutoffs are removed, using the comoving or the physical cutoff schemes can give different flow behavior when an approximation is employed. In the case of inflation, we observe that  the flow obtained with a comoving cutoff function exhibits the trans-Plankian problem. On the other hand, the IR behavior of the flow with the physical cutoff function turns out to be richer.  

It is possible to evaluate the ERG equation, which is an infinite dimensional functional differential equation, for a constant field to obtain a beta functional for the quantum effective potential, which is a system of infinite dimensional ordinary differential equations. For inflation, \eq{s} and \eq{bfp} give the beta functionals of a self-interacting scalar field, in the comoving and the  physical cutoff schemes, respectively. In a time dependent background, one would expect the parameters in the quantum effective action also to depend on time. In applying the LPA, we neglect any plausible time dependencies. Nevertheless, since the whole setup depends on a given time interval $(\eta_i,\eta_f)$, one would expect to see its trace in the equations. For inflation, the  interval is fixed by the number of e-folds $N$, which shows up in the beta functional \eq{s}. One may think that the time dependent parameters of the exact quantum potential are replaced in a way by the time averaged counterparts in the LPA. The situation in the physical cutoff scheme is more interesting. As noted above \eq{bfp}, in that case the time integrals in the left- and right-hand sides of the quantum effective action have the same form leading to \eq{bfp}. This shows that \eq{bfp} is valid for any interval and one can indeed set $\eta_i=-\infty$ and $\eta_f=0$ covering the whole Poincare patch of the de Sitter space (note that \eq{s} is ill defined for $N\to\infty$). Therefore, the ERG flow with the physical cutoff can be used to probe the quantum dynamics of an interacting scalar in the eternal de Sitter space. 

As one of our main results in the present work, we quantitatively determine the ERG flow of the cosmological constant, which is viewed as the constant term in the scalar potential. Starting the flow at some UV scale from a given value of the cosmological constant, we find that the it's value decreases as the cutoff is lowered. Not surprisingly, the largest decrease occurs in the  UV regime, which is of the order of $CH^4$ where $C$ is a very large number depending on the scheme and the UV cutoff, which can presumably be taken as the Planck scale. For the comoving cutoff, the running of the cosmological constant becomes irrelevant at IR, which is also the case for the trivial IR solution  of the physical cutoff scheme \eq{pfs1}. However, there is a special scaling solution to the flow equations in the physical cutoff scheme in which the cosmological constant can be screened by a large negative amount even at IR. These findings are consistent with the expectations that the cosmological constant is relaxed by strong IR quantum effects  \cite{rel1,rel2,rel3,rel4} (or similarly by the backreaction of long wavelength cosmological perturbations \cite{b1,b2,b3}). Although the backreaction effects are neglected in this work, the emergence of screening in a nonperturbative calculation supports  the relaxation mechanism. 

As another interesting finding, we observe that the beta function for the quartic coupling constant changes drastically at IR compared to the flat space behavior. As a result, the coupling constant does not vanish as the IR cutoff is removed, which invalidates the triviality argument. This is a curious IR effect, which emerges due to the presence of spacetime curvature. However, the story is more interesting since in the scaling solution \eq{pfs2} the coupling constant vanishes as the cutoff is removed, although at a different rate compared to the flat space flow. This indicates the existence of a nontrivial phase diagram for the quartically interacting scalar field in de Sitter space. 

It is possible to extend the present work in various different directions. It would be interesting to understand  the backreaction effects along the lines of \cite{ergc1} or \cite{ergc2} to figure out how the screening mechanism affects the gravitational dynamics and the other way around. In addition to the LPA, which is occasionally insufficient in capturing the correct quantum behavior for some cases \cite{rev1}, there are several other approximation methods developed to analyze the ERG flow equations in flat space, some of which can be naturally adopted in the cosmological framework.  For example, considering $O(N)$ scalar models in the large $N$ approximation or applying the systematic derivative expansion as in \cite{mor} are appealing possibilities. Finally, in this paper we approximate the in-in quantum effective action in the form \eq{form}, however it is clear that this approximation is not sufficient to capture all the crucial effects arising from the peculiar properties of the in-in formalism. Besides, in determining the backreaction effects, one should obtain in-in quantum effective {\it field equations}, which can be derived from the in-in action by applying a $\phi^+$ functional derivative and then setting $\phi^+=\phi^-$. It is an important open problem to generalize the ERG formalism for a nonperturbative analysis of the in-in quantum effective field equations. 

\begin{acknowledgments}

I would like to thank the colleagues in the theoretical high energy physics group at McGill University and especially Robert Brandenberger for their hospitality. This work is supported by T\"{U}B\.{I}TAK B\.{I}DEB-2219 grant. 

\end{acknowledgments}

\end{document}